\documentclass[twocolumn,showpacs,preprintnumbers,amsmath,amssymb,superscriptaddress,prb]{revtex4}
\usepackage{dcolumn}
\usepackage{bm}
\usepackage[dvips]{graphicx}
\begin{document}
\title{Modulation of bilayer quantum Hall states by tilted-field-induced subband-Landau-level coupling}
\author{N. Kumada}
\affiliation{NTT Basic Research Laboratories, NTT Corporation, 3-1 Morinosato-Wakamiya, Atsugi, Kanagawa 243-0198, Japan}
\author{K. Iwata}
\affiliation{Research Center for Low Temperature and Materials Sciences, Kyoto University, Kyoto 606-8502, Japan}
\author{K. Tagashira}
\author{Y. Shimoda}
\affiliation{Department of Physics, Tohoku University, Sendai 980-8578, Japan}
\author{K. Muraki}
\affiliation{NTT Basic Research Laboratories, NTT Corporation, 3-1 Morinosato-Wakamiya, Atsugi, Kanagawa 243-0198, Japan}
\author{Y. Hirayama}
\altaffiliation[Present address: ]{Department of Physics, Tohoku University, Sendai 980-8578, Japan}
\affiliation{NTT Basic Research Laboratories, NTT Corporation, 3-1 Morinosato-Wakamiya, Atsugi, Kanagawa 243-0198, Japan}
\affiliation{SORST-JST, 4-1-8 Honmachi, Kawaguchi, Saitama 331-0012, Japan}
\author{A. Sawada}
\affiliation{Research Center for Low Temperature and Materials Sciences, Kyoto University, Kyoto 606-8502, Japan}

\date{Version: \today}

\begin{abstract}
We study effects of tilted magnetic fields on energy levels in a double-quantum-well (DQW) system, focusing on the coupling of subbands and Landau levels (LLs).
The subband-LL coupling induces anticrossings between LLs, manifested directly in the magnetoresistance.
The anticrossing gap becomes larger than the spin splitting at the tilting angle $\theta \sim 20^\circ $ and larger than the cyclotron energy at $\theta \sim 50^\circ $, demonstrating that the subband-LL coupling exerts a strong influence on quantum Hall states even in at a relatively small $\theta $ and plays a dominant role for larger $\theta $.
We also find that when the DQW potential is asymmetric, LL coupling occurs even within a subband.
Calculations including higher-order coupling reproduce the experimental results quantitatively well.
\end{abstract}
\maketitle

\section{Introduction}
\label{introduction}

In a two-dimensional electron system (2DES), an in-plane magnetic field $B_\parallel $ applied parallel to the 2D plane is used to control and investigate quantum Hall (QH) states.
For an ideal 2DES with zero thickness, $B_\parallel $ couples to the system only through the Zeeman energy, while the perpendicular magnetic field $B_\perp $ couples with the orbital degree of freedom as well.
That is, by tilting the magnetic field away from normal to the 2DES while keeping $B_\perp $ fixed, the Zeeman energy is controlled without changing other energies.
This technique has been used to elucidate effects of spins.\cite{Eisenstein,Schmeller}
In a bilayer system, which consists of two parallel 2DESs, the importance of $B_\parallel$ increases because electrons are subjected to Lorentz force and acquire an Aharonov-Bohm (AB) phase \cite{Yang,Hanna} when they tunnel between the two layers.
Lorentz force decreases the tunneling energy gap $\Delta _{\rm SAS}$.\cite{Hu}
This effect, together with the enhancement of the Zeeman energy, is widely used to investigate ground and excited states of bilayer QH states.\cite{Lay,KumadaPRL}
The AB phase affects ground state properties of a coherent bilayer QH state.\cite{Murphy,Ezawabook}

In addition, coupling of the subbands and Landau levels (LLs) in tilted magnetic fields should affect QH states in a system with more than two occupied spatial subbands.
The subband-LL coupling has been manifested as anticrossings between LLs belonging to different subbands in quasi-2D heterojunctions \cite{Schlesinger,Wieck} or parabolic quantum wells.\cite{Ensslin,Gusev,GYu}
This phenomenon provides a new tool to investigate bilayer QH states.
It is therefore important to study the subband-LL coupling in a system with well-defined bilayer potential.
Moreover, since the coupling changes the microscopic properties of wave functions, it affects many-body states as discussed for a collective mode in the coherent bilayer QH state \cite{Yu} or $B_\parallel $-induced reorientation of the stripe phase.\cite{Jungwirthstripe}

In this work, we carried out transport experiments in tilted magnetic fields using a double-quantum-well (DQW) sample, in which the potential symmetry can be tuned by gates deposited on the front and back sides of the sample.
In a perpendicular field, crossings between LLs belonging to different subbands are manifested by a missing or weakening of the QH states at integer fillings.
When the DQW is tilted in the magnetic field, the subband-LL coupling leads to anticrossings.
The anticrossing gap increases with tilting angle $\theta $.
In a symmetric DWQ, it becomes larger than the Zeeman energy at $\theta \sim 20^\circ $ and larger than the cyclotron energy at $\theta \sim 50^\circ $, so that crossings are avoided between LLs that are not coupled to each other.
When the DQW is asymmetric, coupling of LLs in the same subband occur, modifying the anticrossings and the avoided crossings.
We calculated the energy levels from the viewpoint of a single-particle picture and show that they reproduce experiments very well.

This paper is organized as follows.
Section\,\ref{theory} describes a theory by which energy levels in a bilayer system subjected to tilted magnetic fields can be calculated.
In Sec.\,\ref{experiment}, we describe the DQW sample and the experimental setup.
In Sec.\,\ref{result}, the experimental results are presented and compared with the calculation.

\section{Theoretical background}
\label{theory}

\begin{table*}[t]
\caption{Matrix representation of $H$.
$E_\perp ^{N,B}=\hbar \omega _{c\perp }(N+\frac{1}{2})-\frac{\Delta _{\rm BAB}}{2}$ is the eigen energy of $(N,B)$ in a perpendicular field.}
\begin{center}
\renewcommand{\arraystretch}{2}
\begin{tabular}{c|cccc}
& $(N,B)$ & $(N,A)$ & $(N+1,B)$ & $(N+1,A)$ \\
\hline
$(N,B)$ & $E_\perp ^{N,B}+\frac{\hbar \omega_{c\parallel }}{2}\langle B|\frac{z^2}{l_{B\parallel }^2}|B\rangle $ & 0 & $\hbar \omega _{c\perp }\tan \theta \sqrt{\frac{N+1}{2}}\langle B|\frac{z}{l_{B\perp }}|B\rangle $ & $\hbar \omega _{c\perp }\tan \theta \sqrt{\frac{N+1}{2}}\langle B|\frac{z}{l_{B\perp }}|A\rangle $ \\
$(N,A)$ & 0 & $E_\perp ^{N,A}+\frac{\hbar \omega_{c\parallel }}{2}\langle A|\frac{z^2}{l_{B\parallel }^2}|A\rangle $ & $\hbar \omega _{c\perp }\tan \theta \sqrt{\frac{N+1}{2}}\langle A|\frac{z}{l_{B\perp }}|B\rangle $ & $\hbar \omega _{c\perp }\tan \theta \sqrt{\frac{N+1}{2}}\langle A|\frac{z}{l_{B\perp }}|A\rangle $ \\
$(N+1,B)$ & $\hbar \omega _{c\perp }\tan \theta \sqrt{\frac{N+1}{2}}\langle B|\frac{z}{l_{B\perp }}|B\rangle $ & $\hbar \omega _{c\perp }\tan \theta \sqrt{\frac{N+1}{2}}\langle B|\frac{z}{l_{B\perp }}|A\rangle $ & $E_\perp ^{N+1,B}+\frac{\hbar \omega_{c\parallel }}{2}\langle B|\frac{z^2}{l_{B\parallel }^2}|B\rangle $ & 0 \\
$(N+1,A)$ & $\hbar \omega _{c\perp }\tan \theta \sqrt{\frac{N+1}{2}}\langle A|\frac{z}{l_{B\perp }}|B\rangle $ & $\hbar \omega _{c\perp }\tan \theta \sqrt{\frac{N+1}{2}}\langle A|\frac{z}{l_{B\perp }}|A\rangle $ & 0 & $E_\perp ^{N+1,A}+\frac{\hbar \omega_{c\parallel }}{2}\langle A|\frac{z^2}{l_{B\parallel }^2}|A\rangle $
\end{tabular}
\end{center}
\label{matrix}
\end{table*}

In this section, we present a theory describing the effects of an in-plane field on energy levels in a bilayer system.
For the moment, we neglect the spin degree of freedom for simplicity.
Electrons subjected to a tilted magnetic field $(B_\parallel,0,B_\perp )$ in a DQW potential $V(z)$ are described by the Hamiltonian,
\begin{eqnarray}
H=&-&\frac{\hbar ^2}{2m^\ast }\frac{\partial ^2}{\partial x^2}+\frac{e^2B_\perp ^2}{2m^\ast }x^2-\frac{\hbar ^2}{2m^\ast }\frac{\partial ^2}{\partial z^2}+V(z)\nonumber\\&+&\frac{e^2B_\parallel ^2}{2m^\ast }z^2-\frac{e^2B_\parallel B_\perp}{m^\ast}xz,
\label{Hamiltonian}
\end{eqnarray}
where $m^\ast $ is the effective mass.
The first four terms describe electrons in a perpendicular magnetic field, where energy levels are LLs in bonding ($B$) and antibonding ($A$) subbands.
The last two terms of Eq.\,(\ref{Hamiltonian}) stem from the in-plane magnetic field.
The $z^2$ term causes a diamagnetic shift, which merely modifies the energies of the bonding and antibonding states.
The last term is most important for this work: the cross term $xz$ couples the subband and Landau quantizations.

The matrix element of $H$ can be expressed as shown in Table\,\ref{matrix} by using the eigen states in a perpendicular field $(N,\xi )$ as a basis set, where $N$ $(=0,1,...)$ and $\xi $ $(=B,A)$ represent the Landau-orbit and subband indices, respectively.
In a perpendicular magnetic field, the eigen energy of each basis is
\begin{equation}
E_\perp ^{N,B(A)}=\hbar \omega _{c \perp }\left( N+\frac{1}{2}\right) -(+)\frac{\Delta _{\rm BAB}}{2},
\end{equation}
where $\omega _{c\perp }$ is the cyclotron frequency due to $B_\perp $, $\Delta _{\rm BAB}$ is the energy gap between the bonding and antibonding states, which equals $\Delta _{\rm SAS}$ in a symmetric potential and increases with potential asymmetry.
When finite $B_\parallel $ is applied, these diagonal elements undergo a diamagnetic shift of
\begin{equation}
\langle N,\xi |\frac{e^2B_\parallel ^2}{2m^\ast}z^2|N,\xi \rangle =\frac{\hbar \omega _{c \parallel }}{2}\langle \xi |\frac{z^2}{l_{B\parallel }^2}|\xi \rangle ,
\end{equation}
where $\omega _{c\parallel }$ and $l_{B\parallel }=\sqrt{\hbar/eB_\parallel }$ are the cyclotron frequency and magnetic length associated with $B_\parallel $, respectively.
More importantly, off-diagonal elements arise from the $xz$ term of Eq.\,(\ref{Hamiltonian}).
Since $x=(a^\dag +a)/\sqrt{2l_{B\perp}}$  works as the raising $a^\dag $ and lowering $a$ operators of the Landau orbit, it couples LLs when the Landau indices differ by one, where $l_{B\perp }=\sqrt{\hbar/eB_\perp }$ is the magnetic length in $B_\perp $.
As a result, the matrix element between $(N,\xi _i)$ and $(N+1,\xi _j)$ is
\begin{eqnarray}
&&\hspace{-1em}\langle N,\xi _i|\frac{e^2B_\parallel B_\perp}{2m^\ast }xz|N+1,\xi _j\rangle \nonumber \\
&&=\frac{e^2B_\parallel B_\perp}{m^\ast }\langle N|\frac{a^\dag +a}{\sqrt{2l_{B\perp}}}|N+1\rangle \langle \xi _i|z|\xi _j\rangle \nonumber \\
&&=\hbar \omega _{c\perp }\tan \theta \sqrt{\frac{N+1}{2}}\langle \xi _i|\frac{z}{l_{B\perp }}|\xi _j\rangle ,
\label{offdiagonal}
\end{eqnarray}
where $\tan \theta =B_\parallel /B_\perp $.
The spin degree of freedom ($\sigma=\pm 1/2$) can be easily included by adding $\pm \Delta _{\rm Z}/2$ to the diagonal terms, where $\Delta _{\rm Z}$ is the Zeeman energy.

\begin{figure}[b]
\begin{center}
\includegraphics[width=0.45\linewidth]{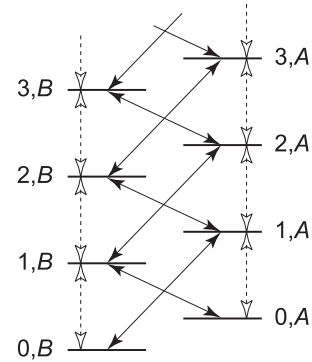}
\caption{Schematic overview of the coupling of energy levels in a bilayer system.
The solid arrows connect levels coupled by the inter-subband-LL coupling.
The dotted arrows represent the intra-subband-LL coupling, which operates when the potential is asymmetric.
}
\label{levelcoupling}
\end{center}
\end{figure}

\begin{figure*}[t]
\begin{center}
\includegraphics[width=\linewidth]{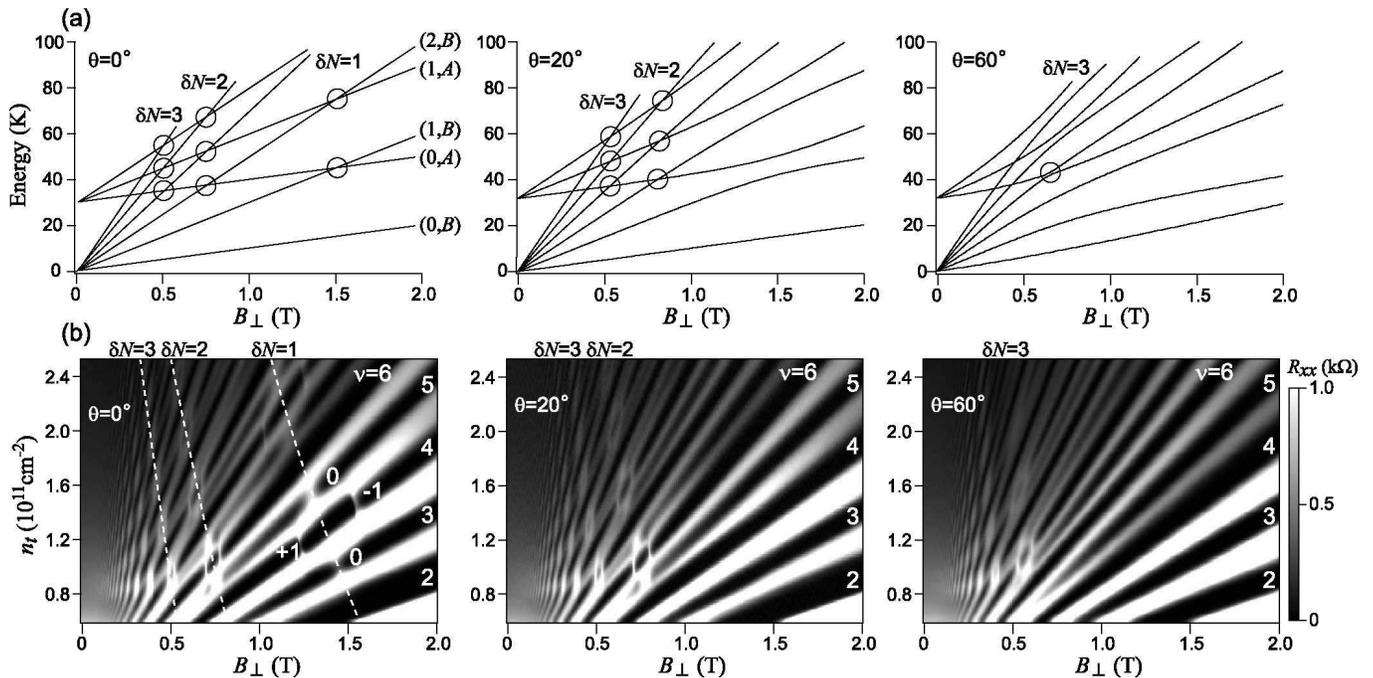}
\caption{(a) Calculated energy levels in a symmetric DQW with $\Delta _{\rm SAS}=31.8$\,K for tilting angles of $\theta =0^\circ $, $20^\circ $, and $60^\circ $.
Though the energy levels are calculated for $N\leq 12$, levels with higher energies are removed for clarity.
The spin splitting is not shown.
(b) Gray-scale plots of $R_{xx}$ as a function of the perpendicular field $B_\perp $ and the total density $n_t$ for the three tilting angles.
The dark regions represent small values of $R_{xx}$.
In the left panel, $\delta \sigma $ for $\delta N=1$ at $\nu =3$-5 is shown.
Dotted lines trace the position of the crossings for $\delta \sigma =0$ at $\delta N=1$-3.
}
\label{overview}
\end{center}
\end{figure*}

It is worth mentioning the effects of the symmetry of the confining potential $V(z)$.
Figure\,\ref{levelcoupling} illustrates the coupling of energy levels in a bilayer system.
When $V(z)$ is symmetric, i.e. $V(z)=V(-z)$, $B$ and $A$ correspond to the symmetric and antisymmetric states, respectively.
Since $z$ is an antisymmetric function, the intra-subband coupling $\langle \xi _i|z|\xi _i \rangle $ (dotted arrows in Fig.\,\ref{levelcoupling}) vanishes because of the symmetry, although the inter-subband coupling $\langle \xi _i|z|\xi _{j\neq i}\rangle $ (solid arrows in Fig.\,\ref{levelcoupling}) is finite.
This symmetry effect causes a selection rule: LLs in different subbands couple only when the difference between the Landau indices is odd as shown by solid arrows in Fig.\,\ref{levelcoupling}.
When $V(z)$ is asymmetric, the intra-subband coupling (dotted arrows in Fig.\,\ref{levelcoupling}) is non-zero, where the odd/even rule becomes less clear.

\section{Experiments}
\label{experiment}

The sample used in this work consists of two 200-\AA-wide GaAs quantum wells separated by a thin 10-\AA-Al$_{0.33}$Ga$_{0.67}$As barrier, processed into a 50-$\mu $m-wide Hall bar with ohmic contacts connecting both layers.\cite{MurakiPhysica}
$\Delta _{\rm SAS}$ decreases from 32 to 23\,K when the total electron density of the two layers is increased from $n_t=0.7$ to $2.3\times 10^{11}$\,cm$^{-2}$.
The low-temperature mobility is 1.2$\times 10^{6}$\,cm$^2/$Vs with the total electron density of the two layers $n_t=2.0\times 10^{11}$\,cm$^{-2}$.
The electron densities in the front layer $n_f$ and the back layer $n_b$ are controlled by the front- and back-gate biases, respectively.
This enables us to control the potential symmetry and $n_t$ independently.
Measurements were performed with the sample mounted in the mixing chamber of a dilution refrigerator with a base temperature of 30\,mK.
By using an ${\it in}$-${\it situ}$ rotator, we can tilt the sample in the magnetic field, which has a maximum strength of 13.5\,T.
Standard low-frequency ac lock-in techniques were used with a current of $I=20$\,nA.

\section{Results and discussion}
\label{result}

In this section, we show experimental data for several tilting angles.
The data are compared with calculated energy levels \cite{calculation} obtained by solving the matrix in Table\,\ref{matrix}.
We are mainly concerned with results for the case where the DQW is kept symmetric (Sec.\,\ref{symmetric}).
Data for asymmetric DQWs are briefly discussed in Sec.\,\ref{asymmetric}.

\subsection{Symmetric potential}
\label{symmetric}

\begin{figure}[b]
\begin{center}
\includegraphics[width=0.95\linewidth]{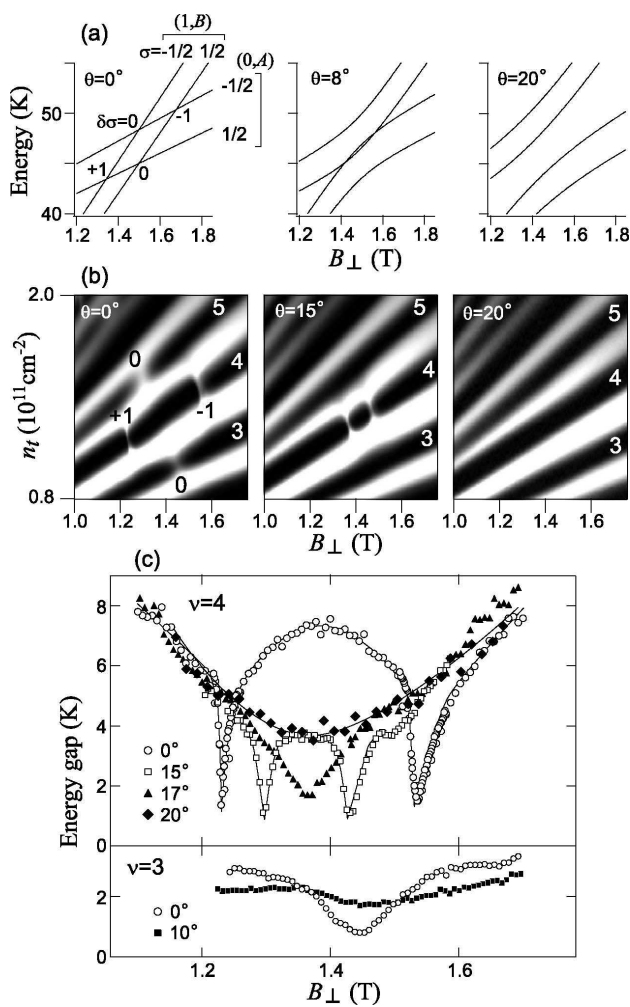}
\caption{(a) Calculated energy levels around $\delta N=1$ crossings at $\nu =3$-5 for $\theta =0^\circ $, $8^\circ $, and $20^\circ $.
(b) Gray-scale plots of $R_{xx}$ for $\theta =0^\circ $, $15^\circ $, and $20^\circ $ around $\delta N=1$ crossings at $\nu =3$-5.
(c) Activation energy gap as a function of $B_\perp $ for fixed filling factors of $\nu =3$ and 4 for several tilting angles.
The lines are guides for the eye.
}
\label{DN1}
\end{center}
\end{figure}

We first show data for symmetric DQW potentials when $n_t$ is scanned while keeping the densities of the two layers equal $(n_f=n_b)$.
Figure\,\ref{overview}(b) shows the magnetoresistance $R_{xx}$ for three tilting angles as a function of $B_\perp $ and $n_t$.
The dark regions represent small values of $R_{xx}$ and thus QH regions.
The left pannel presents the data for the perpendicular field ($\theta =0^\circ $), showing two sets of Landau fans originating from the symmetric and antisymmetric subbands with spin splitting.
Many LL crossings accompanied by an increase in $R_{xx}$ occur when
\begin{equation}
\delta N\times \hbar \omega _{c\perp }+\delta \sigma \times \Delta _{\rm Z}=\Delta _{\rm SAS},
\label{crosspoint}
\end{equation}
where $\delta N$ is the difference between the Landau indices for crossing levels.
Here, $\delta \sigma =0$ ($\pm 1$) identifies crossings between parallel (antiparallel) spin levels.
Note that though $N$ is not a good quantum number in a tilted field, we use the notation $\delta N$ for convenience in identifying LL crossings.
When the sample is tilted at $\theta =20^\circ $, all LL crossings with $\delta N=1$ [those at $\nu =3$-5 around $B_\perp =1.4$\,T and at $\nu =7$-9 around $B_\perp =1.2$\,T in Fig.\,\ref{overview}(b), left] vanish as clearly demonstrated in Fig.\,\ref{overview}(b), center.
At this tilting angle, LL crossings for $\delta N\geq 2$ remain intact.
As $\theta $ is increased further to $\theta =60^\circ $ [Fig.\,\ref{overview}(b), right], the LL crossings for $\delta N=2$ as well as for $\delta N=1$ vanish.

Calculated energy levels based on the subband-LL coupling reproduce the experimental results [Fig.\,\ref{overview}(a)].
In the following, the effects of the subband-LL coupling are discussed quantitatively with experimental results for fine increments of $\theta $ and calculations.

\subsubsection{$\delta N=1$}

We show results for small $\theta $ ($\leq 20^\circ $), focusing on $\delta N=1$ crossings.
Figure\,\ref{DN1}(b) presents the evolution of QH states around the $\delta N=1$ crossings.
In a perpendicular magnetic field, four level crossings with different $\delta \sigma $ occur [Fig.\,\ref{DN1}(b), left].
Note that the sharp features at $\nu =4$ for $\delta \sigma =\pm 1$ are characteristic of the first-order phase transitions associated with easy-axis ferromagnetism.\cite{MurakiPRL,MurakiPhysica}
On the other hand, the broad features at $\nu =3$ and 5 for $\delta \sigma =0$ are due to easy-plane ferromagnetism.\cite{MurakiPRL,MurakiPhysica}
As the sample is tilted at $\theta =15^\circ $, the LL crossings for $\delta \sigma =0$ disappear first [Fig.\,\ref{DN1}(b), center].
At the same time, the magnetic field positions of the two transitions for $\delta \sigma =\pm 1$ approach each other.
Further tilting of the sample ($\theta =20^\circ $) causes them to disappear [Fig.\,\ref{DN1}(b), right].
In Fig.\,\ref{DN1}(c), activation energy gaps at different magnetic fields with fixed filling factors of $\nu =3$ and 4 for several tilting angles are shown, where the activation energy gap $\Delta $ is determined from the temperature dependence of $R_{xx}\propto \exp (-\Delta /2T)$.
For $\nu =4$, the two sharp minima in $\Delta $ first come close to each other $(\theta =15^\circ )$ and then merge $(\theta =17^\circ )$ before turning into a single broad minimum $(\theta =20^\circ )$.
For $\nu =3$, $\Delta $ becomes almost field-independent already at $\theta =10^\circ $.

The experimental results are compared with calculated energy levels [Fig.\,\ref{DN1}(a)].
We note that for the calculation we used enlarged spin splitting $\Delta _{\rm Z}=g^\ast \mu _B(B_\perp ^2+B_\parallel ^2)^{1/2}+2.3B_\perp ^{1/2}$, which effectively incorporates the exchange energy; the value of 2.3 is deduced by the level crossing points for $\delta \sigma =\pm 1$.
In a tilted field, the like-spin levels $(1,B,\sigma )$ and $(0,A,\sigma )$ couple, which causes the anticrossings between these levels with the gap increasing in proportional to $\tan \theta$ [Eq.\,(\ref{offdiagonal})] .
As a result, an energy gap opens in the QH states at $\nu =3$ and 5 [Figs.\,\ref{DN1}(a), center and (c)].
On the other hand, since $\sigma $ is a good quantum number even in a tilted field, the coupling between levels with opposite spins does not occur.
However, as shown in Fig.\,\ref{DN1}(a), right, the $\delta \sigma =\pm 1$ crossings can be avoided when the anticrossing gap exceeds the spin splitting.
We refer to this type of disappearance of the LL crossings as "avoided crossings" to distinguish it from "anticrossings".
The overall similarity between Figs.\,\ref{DN1}(a) and (b) clearly shows that the subband-LL coupling is enough to explain the disappearance of all the LL crossings for $\delta N=1$.
These results demonstrate that the subband-LL coupling exerts strong influence on bilayer QH states even at small $B_{\parallel }$ particularly when LLs belonging to different subbands are close to degeneracy.

\begin{figure}[b]
\begin{center}
\includegraphics[width=0.9\linewidth]{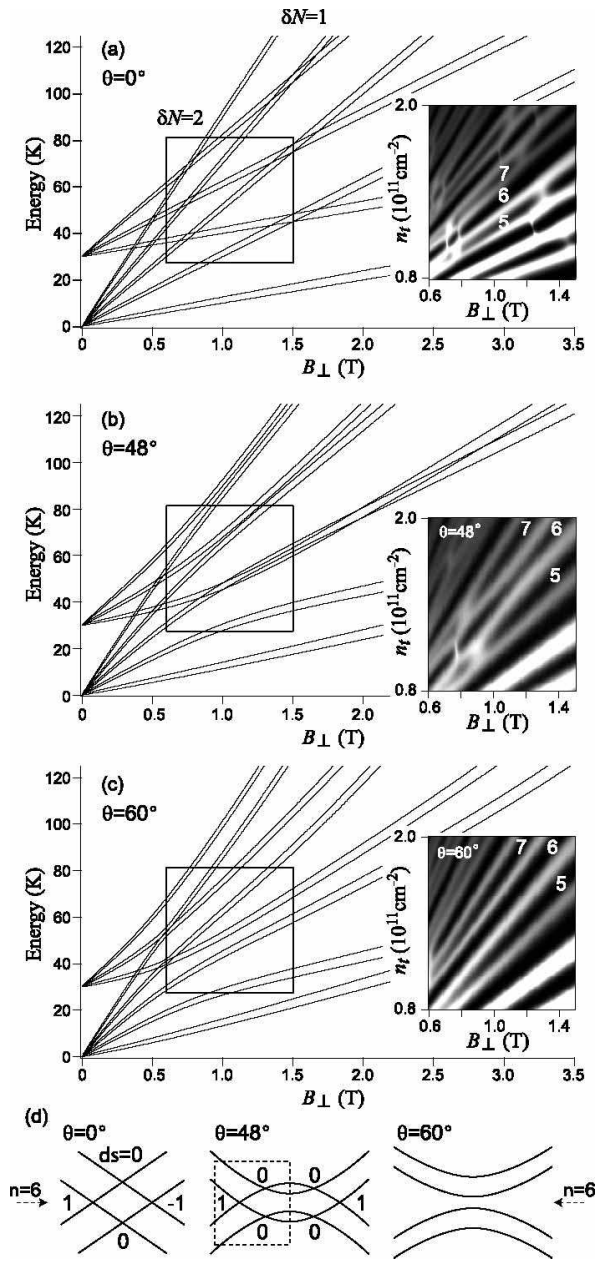}
\caption{Calculated energy levels and Gray-scale plots of $R_{xx}$ around $\delta N=2$ crossings at $\nu =5$-7 crossings for (a) $\theta =0^\circ $, (b) $48^\circ $, and (c) $60^\circ $.
Experimental regions for the gray-scale plots are indicated by the boxes.
(d) Schematic illustration of the evolution of the energy levels.
Dotted lines in the center pannel indicate the experimental region.
}
\label{DN2}
\end{center}
\end{figure}

\begin{figure}[b]
\begin{center}
\includegraphics[width=0.85\linewidth]{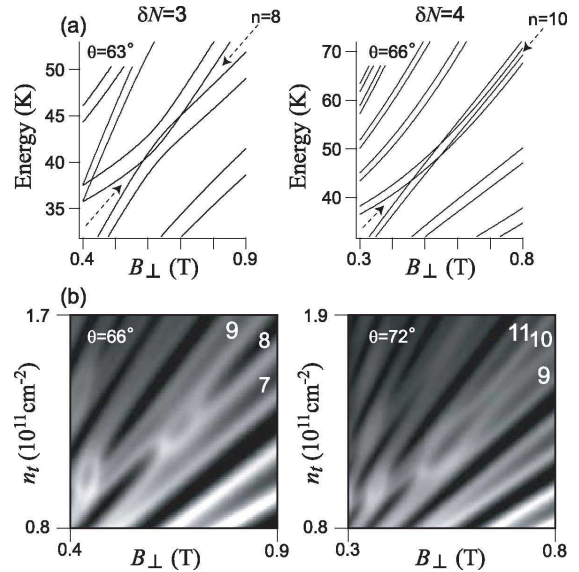}
\caption{(a) Calculated energy levels around $\delta N=3$ crossings at $\nu =7$-9 for $\theta =63^\circ $ (left panel) and $\delta N=4$ crossings at $\nu =9$-11 for $\theta =66^\circ $ (right panel).
(b) Gray-scale plots of $R_{xx}$ around $\delta N=3$ crossings at $\nu =7$-9 for $\theta =66^\circ $ (left panel) and $\delta N=4$ crossings at $\nu =9$-11 for $\theta =72^\circ $ (right panel).
}
\label{DN34}
\end{center}
\end{figure}

\subsubsection{$\delta N=2$}

Next, we show results for larger $\theta $, where LL crossings for $\delta N=2$ disappear.
We focus on those occurring at $\nu =5$-7, which are located around $B_\perp =0.7$\,T at $\theta =0^\circ $ [Fig.\,\ref{DN2}(a)].
Similar to the data for $\delta N=1$, at $\theta =0^\circ $ there are broad (sharp) features for $\delta \sigma =0$ $(\delta \sigma =\pm 1)$ at odd (even) $\nu $.
However, the data at $\theta =48^\circ $ show that the behavior of the $\delta N=2$ crossings is different from that of $\delta N=1$ [Fig.\,\ref{DN2}(b)].
Namely, the $\delta \sigma =-1$ crossing vanishes leaving the $\delta \sigma =1$ crossing at $B=0.8$\,T almost intact.
The absence of the anticrossing for $\delta \sigma =0$ at $\nu =5$ and 7 indicates that there is no subband-LL coupling for $\delta N=2$, consistent with the symmetry arguments.
When the sample is tilted further ($\theta =60^\circ $), all the crossings for $\delta N=2$ disappear [Fig.\,\ref{DN2}(c)].

The behavior of $\delta N=2$ crossings can be explained as the avoided crossings caused by the subband-LL coupling for $\delta N=1$, not for $\delta N=2$.
When the anticrossing gap for $\delta N=1$ becomes larger than $\Delta _{\rm cy}-\Delta _{\rm Z}$, the $\delta \sigma =-1$ crossing for $\delta N=2$ is avoided as schematically shown in Fig.\,\ref{DN2}(d).
Subsequently, as the gap exceeds $\Delta _{\rm cy}+\Delta _{\rm Z}$, LL crossings for $\delta \sigma =0$ and 1 are avoided.
The calculated energy levels agree with experiments on not only the sequence of the avoidance but also field positions of the crossings.
These results show that for large $\theta $, bilayer QH states are strongly modified even though the subband-LL coupling between relevant energy levels does not occur.

\subsubsection{$\delta N=3$ and 4}

We also measured $R_{xx}$ for still larger $\theta $.
Figure\,\ref{DN34}(b) shows a gray-scale plot of $R_{xx}$ around $\delta N=3$ crossings for $\theta =66^\circ $ and that around $\delta N=4$ crossings for $\theta =72^\circ $.
The data for $\delta N=3$ and 4 show similar behavior to that for $\delta N=1$ and 2, respectively.
Namely, the disappearance of the crossings for $\delta N=3$ first occurs for $\delta \sigma =0$ at $\nu =7$ and 9, while that for $\delta N=4$ starts from $\delta \sigma =-1$ at $\nu =10$.
This is evidence of the anticrossing for $\delta N=3$ at $\sigma =0$ and the avoided crossings for $\delta N=4$, confirming the odd/even selection rule.
The calculation explains this behavior quantitatively well [Fig.\,\ref{DN34}(a)].

\begin{figure}[t]
\begin{center}
\includegraphics[width=0.85\linewidth]{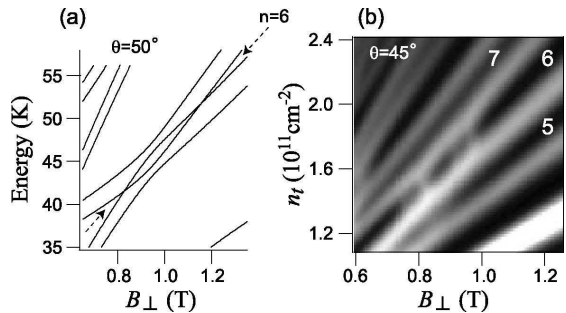}
\caption{(a) Calculated energy levels around $\delta N=2$ crossings at $\nu =5$-7 for $\theta =50^\circ $ in the asymmetric potential with $n_f:n_b=2:1$.
(b) Gray-scale plot of $R_{xx}$ around $\delta N=2$ crossings at $\nu =5$-7 for $\theta =45^\circ $ in the asymmetric potential.
}
\label{asymDN2}
\end{center}
\end{figure}

\subsection{Asymmetric potential}
\label{asymmetric}

Finally, we show data for an asymmetric DQW, where the ratio of the density in the front layer $n_f$ and the back layer $n_b$ is fixed as $n_f:n_b=2:1$.
Figure\,\ref{asymDN2}(b) shows a gray-scale plot of $R_{xx}$ around $\delta N=2$ crossings for $\theta =45^\circ $.
The disappearance of LL crossings first occurs for $\delta \sigma =0$ at $\nu =5$ and 7.
This is contrastive to the data for $\delta N=2$ and 4 in a symmetric potential [Fig.\,\ref{DN2}(b) and right pannel in Fig.\,\ref{DN34}(b)], where the disappearance of LL crossing first occurs at $\delta \sigma =-1$.
This is a clear evidence of the onset of the intra-subband coupling (dotted arrows in Fig.\,\ref{levelcoupling}) and the resultant anticrossing for $\delta N={\rm even}$ [Fig.\,\ref{asymDN2}(a)].
The existence of the intra-subband coupling indicates that even in a system with only one subband, microscopic properties of wave functions are affected by $B_\parallel $.

\section{Conclusions}

In conclusion, we studied effects of tilted magnetic fields on energy levels in a DQW sample by transport measurements and theoretical calculations.
When the sample with symmetric DQW potential is tilted in the magnetic field, LLs belonging to different subbands anticross because of the subband-LL coupling.
The gap of the anticrossing becomes larger than the spin splitting at $\theta \sim 20^\circ $ and the cyclotron energy at $\theta \sim 50^\circ $, causing the avoidance of LLs that are not coupled to each other.
We also carried out experiments in an asymmetric DWQ, showing that the potential symmetry modifies the subband-LL coupling.
Calculated energy levels reproduce the experimental results quantitatively well.
We suggest that since the subband-LL coupling modifies microscopic properties of wave functions, it affects many-body states even in single-layer systems.

\begin{acknowledgments}
The authors are grateful to T. Saku for growing the heterostructures.
\end{acknowledgments}


\begin{thebibliography}{99}
\bibitem{calculation}
For the calculation, $\Delta _{\rm SAS}$ is fixed to 31.8\,K which corresponds to the value at $n_t=1.0\times 10^{11}$\,cm$^{-2}$.
We calculate levels for $N\leq 12$; we confirmed that, in the region we concern, results remain almost unchanged as the number of levels included is increased.
\end{thebibliography}
\end{document}